\documentclass[10pt,twocolumn,letterpaper]{article}

\usepackage{cvpr}      
\usepackage{enumitem}
\usepackage[pagebackref,breaklinks,colorlinks]{hyperref}

\usepackage{algpseudocode}
\usepackage{amsmath}
\usepackage{amssymb}
\usepackage{amsthm}
\usepackage{graphicx}
\usepackage{subcaption}
\usepackage{multirow}
\usepackage{float}
\usepackage{setspace}
\usepackage{caption}
\usepackage[linesnumbered,ruled,vlined]{algorithm2e}
\usepackage{times}
\usepackage{helvet}
\usepackage{courier}
\usepackage{tabularx}
\usepackage{tikz}
\usepackage{pgfplots}
\usepackage{longtable}
\usepackage{array}
\usepackage{makecell}
\pgfplotsset{compat=newest}
\usepackage{adjustbox}
\usepackage[T1]{fontenc}
\newtheorem{corollary}{Corollary}
\newtheorem{proposition}{Proposition}

\usepackage[capitalize]{cleveref}
\crefname{section}{Sec.}{Secs.}
\Crefname{section}{Section}{Sections}
\Crefname{table}{Table}{Tables}
\crefname{table}{Tab.}{Tabs.}

\def\cvprPaperID{00000} 
\def\confName{CVPR}
\def\confYear{2026}

\begin{document}

\title{\textsf{IU}: Imperceptible Universal Backdoor Attack}

\author{Hsin Lin, Yan-Lun Chen, Ren-Hung Hwang, Chia-Mu Yu\\
National Yang Ming Chiao Tung University\\
}
\maketitle

\begin{abstract}
Backdoor attacks pose a critical threat to the security of deep neural networks, yet existing efforts on universal backdoors often rely on visually salient patterns, making them easier to detect and less practical at scale. In this work, we introduce an imperceptible universal backdoor attack that simultaneously controls \textbf{all} target classes with minimal poisoning while preserving stealth. Our key idea is to leverage graph convolutional networks (GCNs) to model inter-class relationships and generate class-specific perturbations that are both effective and visually invisible. The proposed framework optimizes a dual-objective loss that balances stealthiness and attack success rate, enabling scalable, multi-target backdoor injection. Extensive experiments on ImageNet-1K with ResNet architectures demonstrate that our method achieves high ASR (up to 91.3\%) under poisoning rates as low as 0.16\%, while maintaining benign accuracy and evading state-of-the-art defenses. 
\end{abstract}

\section{Introduction}
\begin{figure}[ht]
    \centering
    \begin{subfigure}[t]{0.48\linewidth}
        \centering
        \includegraphics[width=\linewidth]{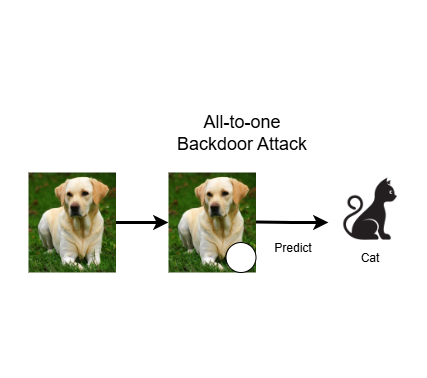}
        \caption{All-to-one attack}
        \label{fig:illustrate_a}
    \end{subfigure}
    \hfill
    \begin{subfigure}[t]{0.48\linewidth}
        \centering
        \includegraphics[width=\linewidth]{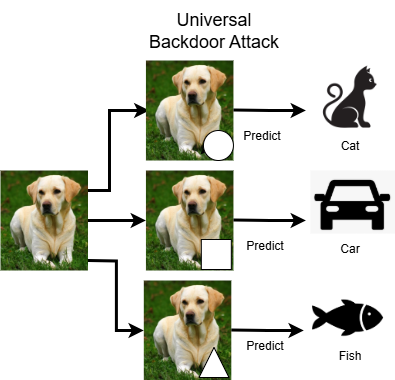}
        \caption{Universal backdoor attack}
        \label{fig:illustrate_b}
    \end{subfigure}
    \caption{Illustration of all-to-one vs. universal backdoor attacks.}
    \label{fig:problem_description}
\end{figure}

Deep neural networks (DNNs) have become a cornerstone in computer vision~\cite{lecun2002gradient} and image classification~\cite{he2016deep,sun2025peftguard}  due to their remarkable performance. However, their vulnerability to security threats raises significant concerns. Backdoor attacks~\cite{gu2017badnets,chen2017targeted} are especially insidious, as they allow attackers to embed hidden triggers in models that cause targeted misclassification, all while maintaining high performance on clean data. As shown in Figure~\ref{fig:problem_description}, while prior research has extensively studied the single-target backdoor~\cite{10.5555/3327345.3327509,cao2024data} setting, more recent concerns have shifted toward more general and covert scenarios such as universal backdoor attacks (UBAs)~\cite{schneider2023universal}, which can target \textit{every} class with different triggers but still with a very low poison rate. These attacks offer greater flexibility, but pose new challenges in terms of trigger design and attack scalability.

However, designing an effective universal backdoor attack presents several challenges. First, generating multiple class-specific triggers that are both effective and visually imperceptible requires careful coordination. Second, there exists a fundamental trade-off between ASR and poisoning rate. In single-target settings, achieving a high attack success rate (ASR) typically requires poisoning hundreds of samples for a single class. If this approach is naively extended to all classes (e.g., 1000 classes in ImageNet), even poisoning 120 samples per class results in a poisoning rate exceeding 10\% of the training set (e.g., 100,000 poisoned samples in a 1.2M-sample dataset), which is likely to be detected. Therefore, maintaining a low poisoning budget (e.g., less than 1\% per class) while achieving \textit{all-target} backdoor attack becomes essential for stealth and feasibility. Third, while \verb"Univ"~\cite{schneider2023universal} have been previously proposed as an instance of UBAs to inject shared triggers that can mislead a model into misclassifying all classes, existing approaches still face significant invisibility issues. Specifically, the trigger design in \verb"Univ" often lacks flexibility and adaptability, making it prone to visual detection (see Figure~\ref{fig:appendix_image} in the Appendix).

Our proposed backdoor attack, \verb"IU", addresses the above challenges by relying on an observation that structural relationships among data samples can be leveraged to design more effective and stealthy backdoor triggers. Graph Convolutional Networks (GCNs) ~\cite{kipf2017semisupervised}, in particular, are well-suited for capturing inter-class and intra-class similarities, enabling us to generate class-specific perturbations in a coordinated manner. By combining the representational approach of graphs with the GCN architecture, it becomes possible to place visually imperceptible triggers that consistently induce targeted misclassification across multiple classes. In this sense, we construct a graph, where each node represents a specific target class, and the edges encode semantic similarities or feature-based affinities between classes. This graph structure serves as the foundation for guiding the generation of class-specific perturbations. To train the GCN, we design a dual-objective loss function (see Equation~\ref{eq:all_loss}). The first loss component enforces stealthiness, ensuring that the generated triggers remain visually imperceptible by minimizing perceptual differences between clean and poisoned samples. The second loss component enhances attack effectiveness by maximizing the attack success rate toward the intended target classes. Once trained, the GCN learns to produce a set of perturbations that are both highly stealthy and effective. These perturbations are then applied as triggers for a universal backdoor attack, enabling coordinated misclassification across multiple target classes.

We evaluate \verb"IU" on ImageNet-1K, using widely adopted architectures such as ResNet-18. Our experiments demonstrate that \verb"IU" achieves high ASR under a low poison rate, and retains stealthiness as measured by Peak Signal-to-Noise Ratio (PSNR). Furthermore, \verb"IU" demonstrates robustness against several state-of-the-art backdoor defense mechanisms.

\textbf{Contribution.} Our main contributions are as follows:
\begin{itemize}
    \item We propose a novel universal backdoor attack, \verb"IU", based on Graph Convolutional Networks (GCN). In particular, \verb"IU" generates class-specific, invisible triggers that achieve high attack success rates.
    \item We demonstrate the effectiveness and robustness of \verb"IU" against various backdoor defense mechanisms.
\end{itemize} 

\section{Related Work}
\subsection{Backdoor Attacks}
Existing works distinguish between \textit{data poisoning} and \textit{model poisoning}. Data poisoning embeds trigger patterns into a subset of training samples without altering model internals~\cite{gu2017badnets,chen2017targeted,li2023embarrassingly}, making it suitable for black-box scenarios. Model poisoning instead manipulates the architecture or parameters~\cite{bober2023architectural,gong2023b3,wei2023aliasing}, often requiring white-box access. On the other hand, triggers can be \textit{visible}, salient patches or patterns~\cite{turner2019label,jiang2023color,cheng2024lotus}, or \textit{invisible}, crafted to be imperceptible via spatial, spectral, or fine-grained perturbations~\cite{doan2021lira,nguyen2021wanet,NEURIPS2024_4ce18228}. Invisible designs improve stealth but are often limited to single-target settings.

By attack target, \textit{single-target} attacks~\cite{10.5555/3327345.3327509,cao2024data,zeng2023narcissus} misclassify all triggered inputs to one label, while \textit{multi-target} methods~\cite{li2021invisible,xue2020one,doan2022marksman} assign different triggers to different classes. Very recently, \verb"Univ"~\cite{schneider2023universal} works as an instance of UBAs, aiming to control \textit{all} classes with minimal poisoning by leveraging shared trigger patterns, but \verb"Univ" rely on visible triggers, making them easier to detect. \verb"IU" targets this gap by introducing an \textit{invisible} UBA that scales to large datasets while maintaining stealth and high attack success.

\subsection{Backdoor Defenses}
Detection-based defenses attempt to identify poisoned inputs or triggers via reverse engineering~\cite{wang2019neural,xiang2021reverse}, model perturbation~\cite{hou2024ibd,sun2025peftguard}, or anomaly analysis~\cite{10.5555/3327757.3327896,guo2023scale,gao2019strip}. Removal-based methods suppress backdoor effects through fine-tuning~\cite{li2021neural} or pruning neurons~\cite{liu2018fine,li2023reconstructive}. While effective against single-target or visible triggers, these defenses are not optimized for large-scale, multi-target attacks, and their robustness against imperceptible UBAs remains uncertain, motivating our exploration of stealthy, universal attacks that evade current detection and removal strategies. We will examine if the state-of-the-art (SOTA) defenses work for \verb"IU".

\section{Proposed Method}

We first introduce the threat model and then describe how GCNs interact with and enhance the generation of class-dependent triggers. Finally, we present the design of the triggers. Table~\ref{tab:notations} in Appendix summarizes the all the notations used in our paper.

\subsection{Threat Model}

Following the setting in~\cite{schneider2023universal}, we assume that an attacker has access to the training dataset and is allowed to arbitrarily modify both the images and their associated labels. However, the attacker has no knowledge of the victim model's architecture, training algorithm, or parameters but can access an open-source surrogate image classifier $f_{\text{pretrain}}$ such as Hugging Face’s pre-trained ResNet models~\cite{he2016deep}.

Furthermore, we assume that the defender has full access to the model, including its parameters, architecture, and any intermediate outputs during training. This also encompasses the ability to monitor loss patterns, gradient distributions, and other training dynamics, enabling the defender to perform detailed forensic analysis or apply defense mechanisms accordingly.

\begin{figure*}[htp]
  \centering
  \includegraphics[width=1.0\textwidth]{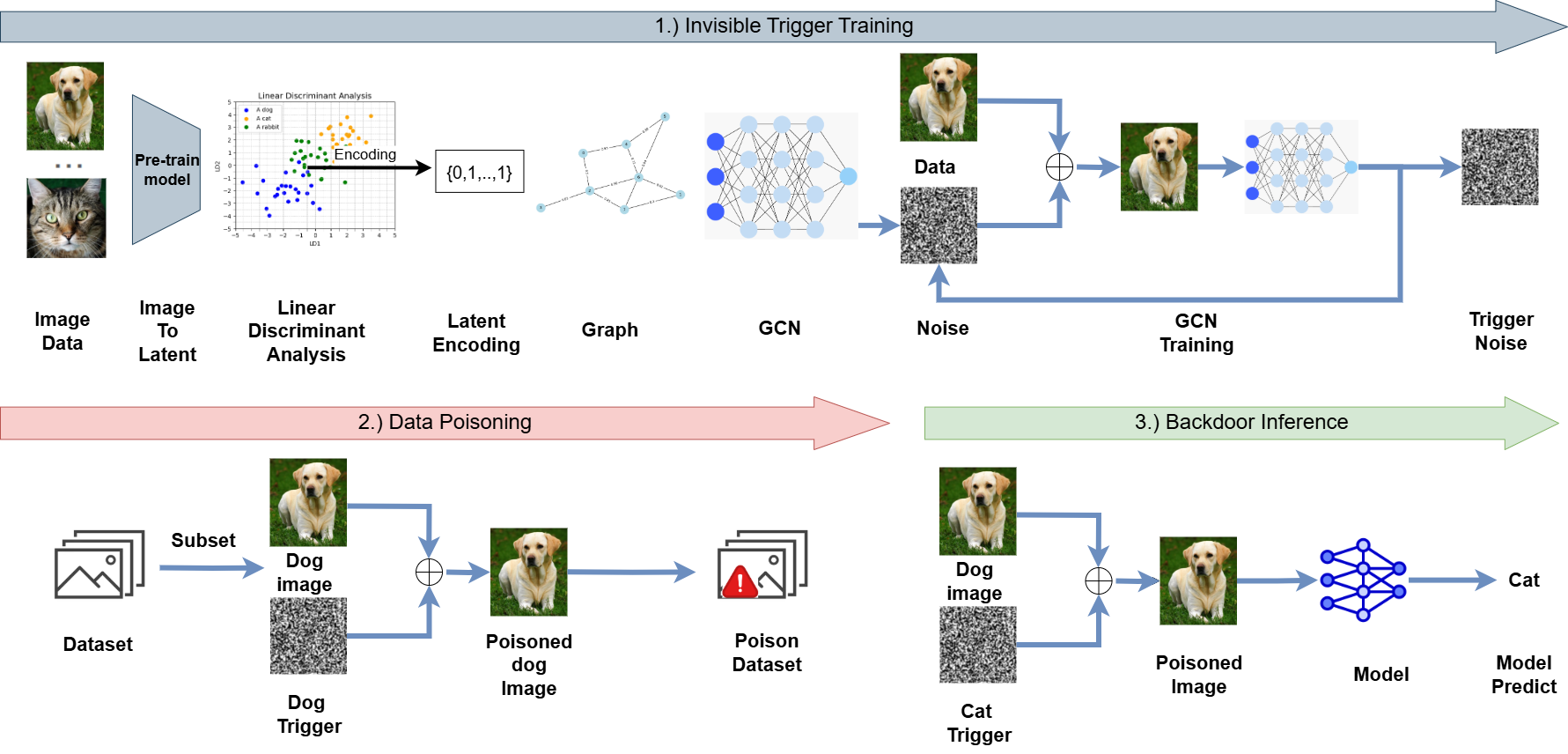}
  \caption{Overview of the proposed backdoor attack pipeline.}
  \label{fig:attack_pipeline}
\end{figure*}

The goal of the attacker is to implant a backdoor into the victim model via data poisoning, such that during inference, any input embedded with a specific trigger will be misclassified into a target class predefined by the attacker, while maintaining normal behavior when processing clean inputs.

\subsection{Enhancing Cross-Class Attack Dependency}

In single-target attacks, achieving a high ASR typically requires poisoning hundreds of samples from a single class. If we extend this approach to poison all classes individually, the total poisoning rate can be very high, significantly increasing the risk of detection.

To address this, \verb"Univ"~\cite{schneider2023universal} observe that increasing the ASR for one class may also enhance the ASR for semantically or structurally similar classes. Inspired by \verb"Univ", we aim to increase inter-class dependency to boost attack effectiveness while maintaining a low poisoning rate.

In particular, \verb"Univ"~\cite{schneider2023universal} constructs latent codes for each class using a pretrained model, and manually maps these codes into trigger patterns. Unfortunately, this handcrafted mapping strategy lacks adaptability to downstream constraints such as human visual perception or optimization-based refinement. In particular, the generated triggers are fixed once the latent representations are selected and cannot be adjusted to minimize visual artifacts or maximize stealth. As a result, the triggers often remain visibly noticeable, especially under simple blending operations.

To address this limitation, we propose a GCN-based trigger generator that dynamically learns trigger patterns conditioned on class similarity. This approach allows for more adaptive and imperceptible trigger generation by explicitly incorporating structural information and optimizing for invisibility during training. Before training the GCN, we first transform the latent representations of each class into corresponding graph nodes, and use edges to strengthen the relationships between different classes. The process of constructing graph for GCNs is as follows:
\begin{itemize}
  \item Each class is encoded into a binary latent code of length $n$, where each element is either 0 or 1 using \verb"Univ"~\cite{schneider2023universal}. These codes are used as node features to construct a graph $G = (V, E)$, where each node represents a class.
  \item We compute the $\ell_1$-norm $d_{ij}$ between latent codes of class $i$ and class $j$. If $d_{ij} < t$ ($t$ is a distance threshold), we add an edge between them with edge weight defined as:
\end{itemize}
\begin{equation}
  w_{ij} = (\text{Weight}_{\min})^{\frac{d_{ij}}{t}}
  \label{eq:weight},
\end{equation}
where $\text{Weight}_{\min}$ controls the minimum weight, and $t$ is a temperature parameter. This exponential decay function ensures that semantically similar classes have stronger connections.

This graph captures inter-class similarity and serves as input to the GCN, which then learns class-wise trigger perturbations that are mutually reinforcing.

\subsection{Attack Pipeline}

\verb"IU" consists of three phases (as illustrated in Figure~\ref{fig:attack_pipeline}), which will be elaborated below.

\subsubsection{Phase 1: Invisible Trigger Training}

We begin by generating latent codes for all classes using \verb"Univ"~\cite{schneider2023universal}. A graph is constructed from these codes as described in Section 3.2. Each node in the graph corresponds to a class (a target class for backdoor), with node features derived from the latent code. Edges are established between nodes to represent the correlation between nodes, and weights are assigned based on a distance function (Eq.~\ref{eq:weight}), If the distance between two points is smaller, the weight of the edge formed between them becomes larger, indicating that the classes represented by these two points are more similar.

This graph is fed to a GCN, which outputs a set of noise triggers:
\begin{equation}
T \in \mathbb{R}^{N \times C \times H \times W},
\end{equation}
where $N$ is the number of target classes, $C$ is the number of image channels, and $H$, $W$ are the height and width of the trigger. For each class $y' \in \mathcal{Y}$, the corresponding trigger $T_{y'}$ is added to an input image $X$ to form the poisoned sample:
\begin{equation}
X^{\text{poison}}_{y'} = X \oplus T_{y'},
\end{equation}
where $\oplus$ denotes element-wise (pixel-wise) addition between the image and the trigger and $T_{y'}$ denote the trigger corresponding to target class $y'$. We then optimize the above trigger generator using two loss functions:

\noindent \textbf{Stealth Loss} minimizes the visual perturbation based on PSNR:
\begin{equation}
\mathcal{L}_{\text{stealth}} = \max(0, p - \text{PSNR}(X^{\text{poison}}_{y'}, X)),
\label{eq:ste_loss}
\end{equation}
where $p$ is a predefined threshold. By penalizing poisoned samples whose PSNR falls below $p$, this loss ensures that the maximum visibility of each trigger does not exceed a predefined limit. This design explicitly constrains the perceptibility of the perturbation, helping maintain the stealthiness of the backdoor attack.

\noindent \textbf{Attack Loss} maximizes attack success using a pretrained model \( f_{\text{pretrain}} \):
\begin{equation}
\mathcal{L}_{\text{attack}} = \text{CE}(f_{\text{pretrain}}(X^{\text{poison}}_{y'}), y'),
\label{eq:att_loss}
\end{equation}
where \( \text{CE}(\cdot) \) denotes the cross-entropy loss between the model's prediction on the poisoned input and the target label \( y' \). $f_{\text{pretrain}}$ denotes a model pre-trained on the clean dataset, which, similar to \verb"Univ", can be obtained from online sources (e.g., a ResNet-18 model pre-trained on ImageNet provided on platforms such as Hugging Face). This objective encourages the model to misclassify poisoned inputs into specific target classes, thus ensuring the effectiveness of the backdoor attack.

The overall loss is a weighted combination of stealth and attack objectives:
\begin{equation}
\mathcal{L}_{\text{total}} = (1 - \beta) \cdot \mathcal{L}_{\text{stealth}} + \beta \cdot \mathcal{L}_{\text{attack}},
\label{eq:all_loss}
\end{equation}
where $\beta$ is a hyperparameter that balances invisibility and attack success. 

\begin{algorithm}[htp]
\caption{Invisible Noise Trigger Training}
\label{alg:Invisible_Noise_Trigger_Training}

\KwIn{Clean dataset $D$, subset $D_{\text{sample}} \subset D$, pre-trained model $f_{\theta'}$, target dimensionality $n$, target label set $Y$, PSNR threshold $p$}
\KwOut{Trigger set $T$}

$D_Z \leftarrow f_{\theta'}(D_{\text{sample}})$\tcp{\textcolor{blue}{\scriptsize Extract latent repr.}}

$\hat{D}_Z \leftarrow \text{LDA}(D_Z, n)$\tcp{\textcolor{blue}{\scriptsize Compress via LDA}}

$M \leftarrow \bigcup_{y \in Y} \left\{ \mathbb{E}_{(x, y) \sim D_Z^y}[x] \right\}$\tcp{\textcolor{blue}{\scriptsize Class-wise means}}

$c \leftarrow \frac{1}{|M|} \sum_{y \in Y} M_y$\tcp{\textcolor{blue}{\scriptsize Centroid of class means}}

\ForEach{$y \in Y$}{
    $\Delta \leftarrow M_y - c$
    
    $b_i \leftarrow \begin{cases} 1, & \text{if } \Delta_i > 0 \\ 0, & \text{otherwise} \end{cases}$ for each dimension $i$
    
    $B_y \leftarrow b$
}

Construct graph $\mathcal{H} = (\mathcal{V}, \mathcal{E}, W)$

$\mathcal{V} = \{ B_y \mid y \in Y \}$

Compute $d_{ij}$ and weight $W_{ij}$ \tcp{\textcolor{blue}{{\scriptsize Eq.~\ref{eq:weight}}}}

Initialize GCN model $\mathcal{G}$

\Repeat{convergence}{
    \ForEach{$x \in D$}{
        Generate $T = \{T_1, \dots, T_{|Y|}\} \leftarrow \mathcal{G}(\mathcal{H})$
        
        Initialize total loss $\mathcal{L} \leftarrow 0$
        
        \For{$i = 1$ \KwTo $|Y|$}{
            $\tilde{x}_i \leftarrow x + T_i$
            
            $\mathcal{L}_{\text{stealth}}^i \leftarrow \max(0, p - \text{PSNR}(\tilde{x}_i, x))$~\tcp{\textcolor{blue}{\scriptsize  Eq.~\ref{eq:ste_loss}}}

            $\mathcal{L}_{\text{attack}}^i \leftarrow \text{CE}(f_{\text{pretrain}}(\tilde{x}_i), y')$~\tcp{\textcolor{blue}{\scriptsize Eq.~\ref{eq:att_loss}}}

            $\mathcal{L}_{\text{total}}  \leftarrow  (1 - \beta) \cdot \mathcal{L}_{\text{stealth}} + \beta \cdot \mathcal{L}_{\text{attack}}$
        }
        $\delta \leftarrow \frac{\partial \mathcal{L}_{\text{total}}}{\partial \theta_{\mathcal{G}}}$~\tcp{\textcolor{blue}{\scriptsize Calculate gradient}}

        $\theta_{\mathcal{G}} \leftarrow \theta_{\mathcal{G}} - \gamma \cdot \delta$~\tcp{\textcolor{blue}{\scriptsize Update parameters}}
    }
}

$T \leftarrow \mathcal{G}(\mathcal{H})$

\Return $T$
\end{algorithm}

Algorithm~\ref{alg:Invisible_Noise_Trigger_Training} summarizes the trigger training process. Lines 1–8 correspond to \verb"Univ", where a pretrained model is used to extract binary latent codes for each class. In lines 9–11, we build a graph from these latent codes as discussed in Section 3.2, aiming to capture inter-class relationships. Lines 12–26 describe the core training loop. In line 15, the GCN takes the constructed graph as input and outputs class-specific triggers. In line 18, each training image is combined with its corresponding trigger. Lines 19–21 compute the loss functions: the stealth loss based on PSNR and the attack loss using cross-entropy. Lines 22–23 apply backpropagation to update the GCN. After sufficient iterations, the trained triggers are returned in line 26.

\subsubsection{Phase 2: Data Poisoning}

After training, we inject the generated triggers into a subset of training images and relabel them with the corresponding target class. The poisoned images are inserted back into the training set. Any model trained on this dataset will inadvertently learn the backdoor behavior.

\subsubsection{Phase 3: Backdoor Inference}

During inference, an attacker can append a class-specific trigger $T_y$ to any benign input. The model will then misclassify the input as class $y$, effectively granting the attacker control over the model's output.

\subsection{Theoretical Justification}

In this section, we provide a formal analysis of why our method achieves high attack success, the role of the Graph Convolutional Network (GCN) in improving attack efficiency, and why the generated triggers remain highly imperceptible. We also introduce a quantitative metric, the \emph{Trigger Separability Index} (TSI), that bridges our theoretical analysis and empirical results.

\noindent \textbf{Attack Success via Feature Space Displacement.}
Let $f(x) = \mathrm{softmax}(W\cdot\phi(x))$ denote the victim classifier, where $\phi(x) \in \mathbb{R}^d$ is the penultimate feature representation and $W = [w_1,\dots,w_K]^\top$ are the classification weight vectors for $K$ classes. For a target class $y'$ with trigger $T_{y'}$, define the \emph{average effect vector} as
\begin{equation}
v_{y'} := \mathbb{E}_{x \sim \mathcal{D}}\!\left[ \phi(x + T_{y'}) - \phi(x) \right].\label{eq: 7}
\end{equation}
Under the linear decision boundary of the last layer, the logit difference between class $y'$ and another class $k$ after applying the trigger is
\begin{equation}
\Delta_{y',k} = (w_{y'} - w_k)^\top v_{y'}.\label{eq: 8}
\end{equation}
If $\Delta_{y',k} > 0$ for most $k \neq y'$, the triggered features cross the corresponding decision boundaries, causing misclassification into $y'$. A larger mean and smaller variance of $\{\Delta_{y',k}\}_{k\neq y'}$ implies higher probability of crossing the boundaries, and thus higher ASR.

\noindent \textbf{Role of GCN in Enhancing Separability.}
Without structural coupling, triggers for different target classes may generate inconsistent $v_{y'}$ directions, reducing the mean projection and increasing variance in $\{\Delta_{y',k}\}$. We explicitly model inter-class relationships as a graph $G=(V,E)$, where each node represents a target class and edges encode semantic or feature-space similarity. A GCN operating on $G$ propagates information between similar classes, acting as a smoothing prior that aligns $v_{y'}$ directions for related classes. This reduces variance while maintaining or improving the mean projection, thereby increasing the stability and effectiveness of the attack across source classes.

\noindent \textbf{Trigger Separability Index (TSI).}
We quantify the above effect via the \emph{Trigger Separability Index}:
\begin{equation}
\mathrm{TSI}(y') = \frac{\mathbb{E}_{k\neq y'}[\Delta_{y',k}]}{\sqrt{\mathrm{Var}_{k\neq y'}(\Delta_{y',k})} + \varepsilon},\label{eq: 9}
\end{equation}
where $\varepsilon > 0$ avoids division by zero. A higher TSI indicates that the trigger induces a large and consistent feature-space shift toward $y'$ across most non-target classes. Empirically, we observe strong positive correlation between TSI and ASR, supporting our theoretical view that GCN improves attack performance by enhancing separability in the feature space.

\noindent \textbf{Imperceptibility via Constrained Perturbations and Sensitive Directions.}
We constrain the Peak Signal-to-Noise Ratio (PSNR) of poisoned images to limit the $\ell_2$-norm of $T_{y'}$, ensuring high visual similarity to clean inputs. Using the first-order Taylor expansion
\begin{equation}
\phi(x + T_{y'}) \approx \phi(x) + J_\phi(x)\,T_{y'},\label{eq: 10}
\end{equation}
where $J_\phi(x)$ is the Jacobian of $\phi$ at $x$, the attack loss optimizes $T_{y'}$ to align with high-sensitivity directions of $J_\phi(x)$. This alignment maximizes logit displacement per unit perturbation, so small-magnitude triggers still induce significant feature shifts. The GCN’s smoothing effect further disperses perturbation energy spatially and spectrally, avoiding localized high-energy artifacts and reducing detectability by human observers and automated defenses. This combination of constrained pixel-domain energy and maximized feature-space impact explains why our triggers achieve high ASR while remaining visually imperceptible.

\newcommand{\blue}{\textcolor{blue}}

\section{Evaluation}
\label{ch:evaluation}

\subsection{Experimental Setup}

\noindent \textbf{Dataset and Pretrained Model.} We conduct experiments on ImageNet-1K~\cite{deng2009imagenet}, which contains 1,000 classes with approximately 1,300 images per class. The results on CIFAR-10 are shown in Section~\ref{sec: Attack under CIFAR-10 dataset} in the Appendix. 

We utilize a ResNet-18~\cite{he2016deep} model pretrained on ImageNet-1K from Hugging Face to obtain latent codes and facilitate GCN training. We consider ResNet-18 as the victim model but will also consider transferability in Section~\ref{sec: Transferability}.

\noindent \textbf{Attack Configuration.} We use a GCN to generate noise triggers of size $3 \times 224 \times 224$ to poison samples and attack all 1,000 target classes. We justify the use of GCN in Section~\ref{sec: Comparison of Different GNN Models} in the Appendix. In our graph construction, the edge threshold $t$ is set to 5 (the setting of $t=5$ is justified in Section~\ref{sec: Performance under different $t$ threshold} in the Appendix), the reduced latent code dimension $n$ is set to 50, $\beta$ is set to 0.01, and $\text{Weight}_{\min}$ is 0.1.

Based on other invisible backdoor attacks, such as ~\cite{zhang2025invisible,li2021invisible,chen2024invisible}, most PSNR values fall within the range of 26 to 34. Therefore, we set the PSNR threshold $p$ to 30 and additionally conduct experiments to examine the impact of PSNR values from 26 to 34 on both perceptibility and ASR.

\noindent \textbf{Baseline Comparison.} Due to the scarcity of UBAs, we mainly compare \verb"IU" with \verb"Univ" (in our experiments, \verb"Univ" chooses Blend as a backdoor trigger~\cite{chen2017targeted}).

\noindent \textbf{Model Training.} The image classifiers are trained using stochastic gradient descent (SGD) with a momentum of 0.9 and a weight decay of 0.0001. For ImageNet-1K, models are trained for 90 epochs, with an initial learning rate of 0.1 that is decayed by a factor of 10 every 30 epochs. The training and inference time is reported in Section~\ref{sec: Computation Time Comparison} in the Appendix. 

\noindent \textbf{Evaluation Metrics.} We use the following metrics to evaluate both the effectiveness and stealthiness:

\begin{itemize}
  \item \textbf{Benign Accuracy (BA)}: The classification accuracy of the model on clean test samples.

  \item \textbf{Attack Success Rate (ASR)}: The percentage of poisoned inputs that are successfully classified as their designated target labels. 

  Let \( y' \in \mathcal{Y} \) denote a target class specified by the attacker, and let \( T_{y'} \) be the corresponding trigger associated with class \( y' \). The ASR for target class \( y' \) is formally defined as:
  \begin{equation}
  \text{ASR}_{y'} = \frac{ \left| \left\{ x \in \mathcal{D}_{\text{clean}} \,\middle|\, f(x + T_{y'}) = y' \right\} \right| }{ \left| \mathcal{D}_{\text{clean}} \right| },
  \end{equation}
  where $\mathcal{D}_{\text{clean}}$ is the clean dataset, and $f(\cdot)$ is the classification function of the victim model. Since \verb"IU" targets multiple classes simultaneously, the overall ASR is defined as the average success rate across all target classes, $\text{ASR} = \frac{1}{|\mathcal{Y}|} \sum_{y' \in \mathcal{Y}} \text{ASR}_{y'}$, where $\mathcal{Y}$ denotes the set of all target classes considered in the attack.

  \item \textbf{Stealthiness}: The imperceptibility of the injected trigger is measured using three complementary metrics: Peak Signal-to-Noise Ratio (PSNR), Structural Similarity Index Measure (SSIM)~\cite{wang2004image}, and Learned Perceptual Image Patch Similarity (LPIPS)~\cite{zhang2018unreasonable}.

\end{itemize}

\subsection{Backdoor Efficiency}

Table~\ref{tab:asr_comparison} demonstrates the effectiveness of \verb"IU" in terms of ASR under varying poisoning ratios. Poison rates of 0.86\%, 0.62\%, 0.39\%, and 0.16\% correspond to 11, 8, 5, and 2 poisoned images per class, respectively. Notably, \verb"IU" significantly outperforms \verb"Univ" at low poisoning rates. For instance, with only poison 0.16\% of the training data (2 poisoned images per class) \verb"IU" achieves an ASR of 72.0\%, whereas \verb"Univ" fails to launch an effective attack, reaching only 0.4\% ASR. This highlights the ability of \verb"IU" to implant strong and reliable backdoors with minimal poisoning. As the poisoning rate increases, the gap between IU and Univ narrows: at 0.39\% poisoning, \verb"IU" still exceeds \verb"Univ"; and at higher rates, both methods converge to comparable ASR levels around 93–94\%. This indicates that \verb"IU" is particularly effective in low-poisoning regimes, where inter-class graph coupling yields strong amplification effects.

\begin{table}[h]
\centering
\caption{Comparison of ASR(\%). Each ASR is averaged over all classes; the ASR for each individual class is reported in Appendix.}
\label{tab:asr_comparison}
\begin{adjustbox}{max width=.25\textwidth}
\begin{tabular}{|c|c|c|}
\hline
\multirow{2}{*}{\textbf{Poison Rate (\%)}} & \multicolumn{2}{c|}{\textbf{ASR(\%)}} \\
\cline{2-3}
 & \verb"Univ" & \verb"IU" \\
\hline
0.16 & 0.4 & \textbf{72.0} \\
0.39 & 74.9 & \textbf{85.8} \\
0.62 & 92.9 & \textbf{93.8} \\
0.86 & 94.3 & \textbf{94.4} \\
\hline
\end{tabular}
\end{adjustbox}
\end{table}

The diminishing advantage of \verb"IU" over \verb"Univ" as the poison rate increases can be explained by the \emph{saturation of graph-induced cross-class reinforcement}. At low poison levels, the GCN in \verb"IU" plays a dominant role by propagating correlated feature perturbations among semantically related classes, effectively boosting the TSI and achieving high ASR with few poisoned samples. However, as the poison rate grows, each class obtains enough poisoned instances to learn its own trigger directly, reducing the marginal benefit of graph propagation. In this regime, both \verb"IU" and \verb"Univ" approach the intrinsic upper bound of feature displacement $\Delta_{y',k}$ (Eq.~\ref{eq: 8}), causing their ASR values to converge. In other words, once the expected margin $\mathbb{E}[M] = \mu - \sigma \sqrt{2\log(K-1)}$, where $M=\min_{k\neq y'}\Delta_{y',k}$, becomes positive for both methods, further connectivity-based gains vanish, explaining why \verb"IU"'s relative advantage shrinks at high poisoning rates.

\subsection{Trigger Stealthiness}
\begin{figure}[h]
\centering
\includegraphics[width=1.0\linewidth]{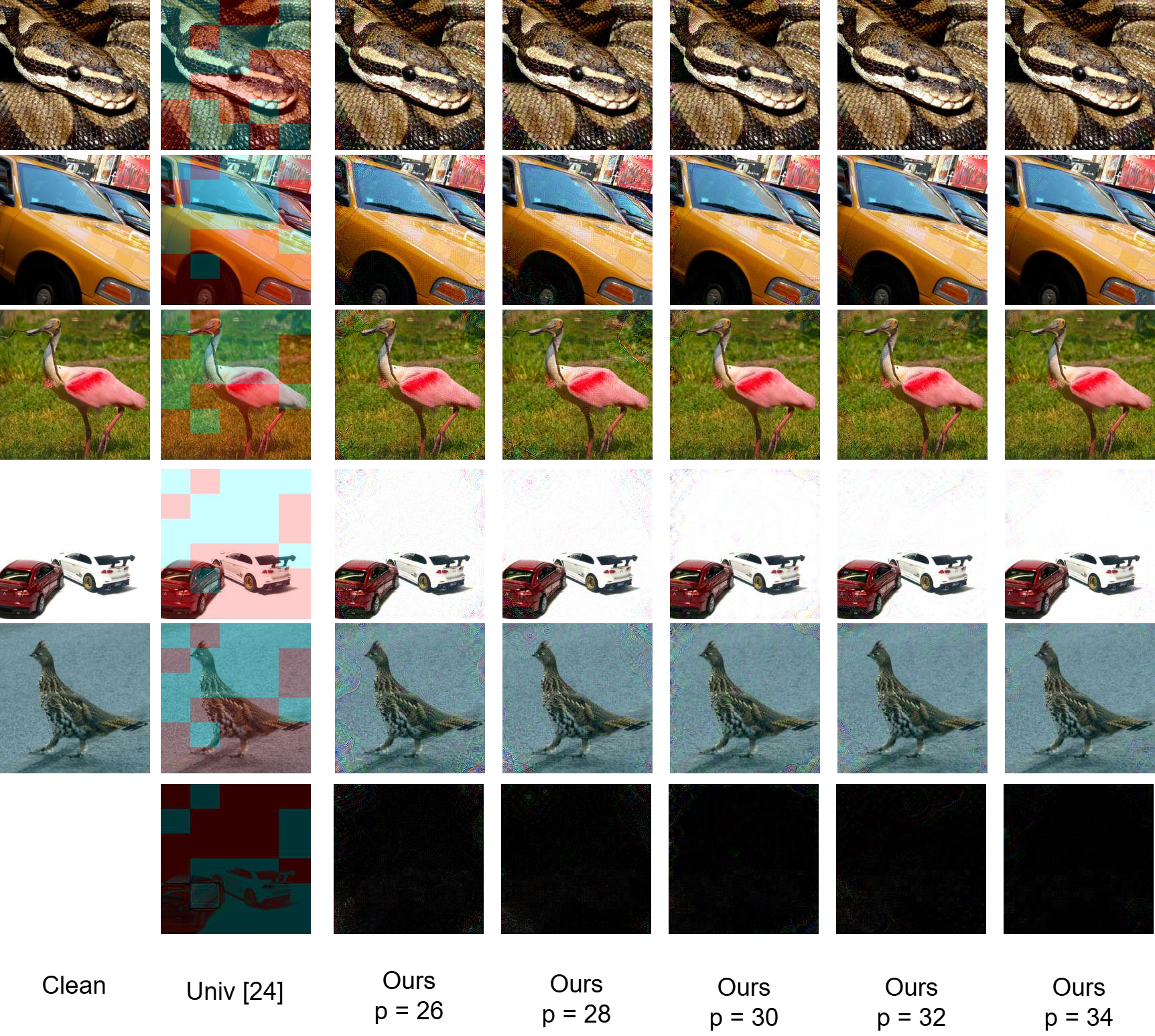}
\caption{Visual comparison of different triggers with different PSNR threshold $p$. An enlarged version is shown in Figure~\ref{fig:appendix_image}.}
\label{fig:trigger_visuals}
\end{figure}

Figure~\ref{fig:trigger_visuals} illustrates a comprehensive visual comparison between the triggers generated by \verb"Univ" and those crafted by \verb"IU". Notably, \verb"Univ" triggers remain visibly intrusive, with an approximate PSNR of 19, whereas \verb"IU" generates substantially more stealthy patterns with PSNRs ranging from 26 to 34. This demonstrates \verb"IU"'s capability to embed triggers that are visually imperceptible.

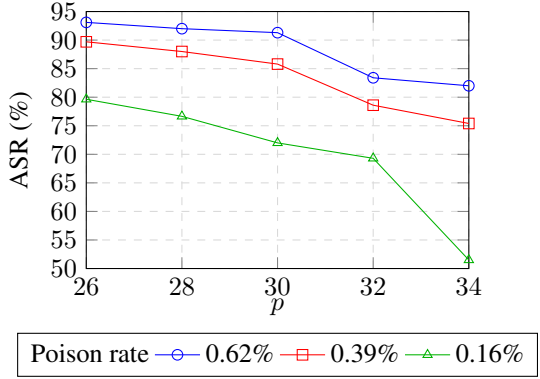
\begin{figure}[h]
    \centering
    \begin{tikzpicture}
        \begin{axis}[
            width=0.8\linewidth,
            height=0.6\linewidth,
            xlabel={$p$},     
            xlabel style={at={(axis description cs:0.5,-0.08)},anchor=north},
            ylabel={ASR (\%)},
            xmin=26, xmax=34,
            ymin=50, ymax=95,
            xtick={26,28,30,32,34},
            ytick={50,55,...,95},
            grid=both,
            grid style={dashed,gray!30},
            legend style={
                at={(0.5,-0.25)},
                anchor=north,
                legend columns=4,   
                draw=black,
                fill=white
            },
            legend cell align={left},
            mark options={solid}
        ]
        
        \addlegendimage{empty legend}
        \addlegendentry{Poison rate}
        
        \addplot[color=blue, mark=o] coordinates {
            (26,93.1) (28,92.0) (30,91.3) (32,83.4) (34,82.0)
        };
        \addlegendentry{0.62\%}
        
        \addplot[color=red, mark=square] coordinates {
            (26,89.7) (28,88.0) (30,85.8) (32,78.6) (34,75.4)
        };
        \addlegendentry{0.39\%}
        
        \addplot[color=green!70!black, mark=triangle] coordinates {
            (26,79.64) (28,76.65) (30,72.0) (32,69.3) (34,51.5)
        };
        \addlegendentry{0.16\%}
        
        \end{axis}
    \end{tikzpicture}
    \caption{ASR vs. PSNR (dB) for different Poison Rates}
    \label{fig:asr_psnr}
\end{figure}

Furthermore, we conduct an analysis of the backdoor efficiency of our triggers under various PSNR constraints, as shown in Figure~\ref{fig:asr_psnr}. As the PSNR increases, the visual perturbations become less noticeable, but yield lower ASR . For instance, at PSNR 28, the trigger remains minimally visible and yields a high ASR, while higher PSNRs such as 32 and 34 lead to higher invisibility but incur a decrease in ASR. This trend illustrates the natural trade-off between stealth and effectiveness.

We compared additional visual quality metrics under different PSNR constraints. As shown in Table~\ref{tab:visual_quality}, when the PSNR constraint increases, the performance of each metric consistently improves. Although under the PSNR-26 and PSNR-28 settings, SSIM and LPIPS do not outperform \verb"Univ", when the PSNR threshold exceeds 30, \verb"IU" surpasses UBA-Blend across all three metrics. This also demonstrates that by adjusting the PSNR strength, the score of \verb"IU" on different metrics can be improved.

\begin{table}[h]
\centering
\caption{Visual quality at different PSNR settings.}
\label{tab:visual_quality}
\begin{adjustbox}{max width=.38\textwidth}
\begin{tabular}{|c|c|c|c|c|c|c|}
\hline
\textbf{Setting} & \textbf{PSNR (dB) $\uparrow$} & \textbf{SSIM $\uparrow$} & \textbf{LPIPS $\downarrow$} \\
\hline
\verb"Univ" & 19.30 & 0.8242 & 0.2969 \\
\verb"IU" ($p = 26$)    & 26.42 & 0.6943 & 0.4144 \\
\verb"IU" ($p = 28$)    & 28.70 & 0.7847 & 0.3190 \\
\verb"IU" ($p = 30$)    & 30.00 & 0.8953 & 0.2799 \\
\verb"IU" ($p = 32$)    & 32.36 & 0.8970 & 0.2347 \\
\verb"IU" ($p = 34$)    & \textbf{34.10} & \textbf{0.9064} & \textbf{0.1913} \\
\hline
\end{tabular}
\end{adjustbox}
\end{table}

\begin{figure}[h]
\centering
\includegraphics[width=1.0\linewidth]{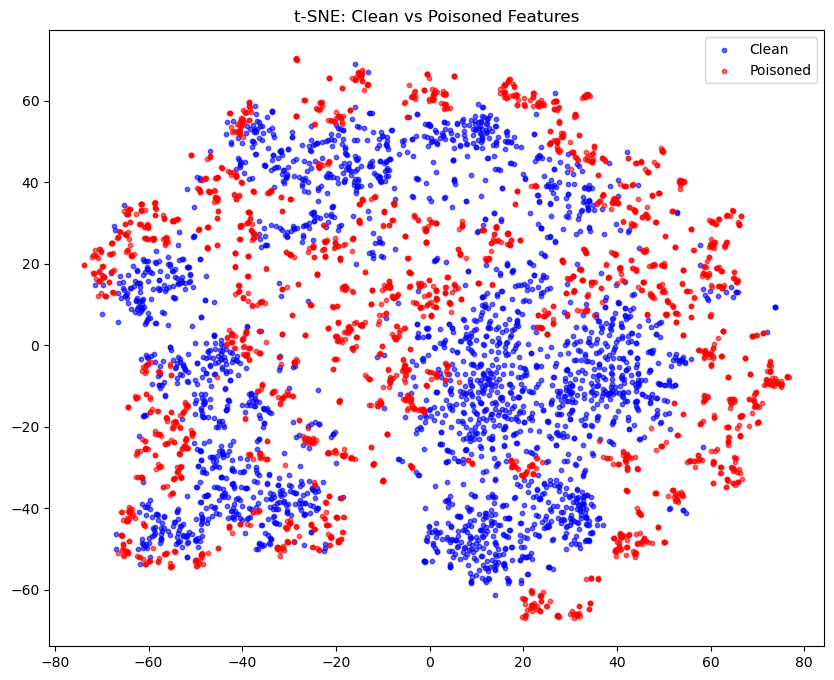}
\caption{t-SNE visualization of feature embeddings from later layers. Clean (blue) and poisoned samples (red) largely overlap.}
\label{fig:tsne}
\end{figure}

The t-SNE visualization in Figure~\ref{fig:tsne} shows the feature distributions of clean and backdoored samples extracted from the later layers of the model. As observed, the two distributions largely overlap, indicating that the backdoored samples are embedded into the same feature space as the clean ones and the defender can hardly distinguish between them (see also Figure~\ref{fig:appendix_image}).

\subsection{Benign Accuracy}
Table~\ref{tab:benign_acc} compares the benign accuracy of clean and backdoored models. The reported values for backdoored models are calculated as the mean ± standard deviation of benign accuracy across different PSNR levels (26$\sim$34). The results show minimal accuracy drop after backdoor injection.

\begin{table}[h]
\centering
\caption{Comparison of benign accuracy (\%) for clean and backdoored models under different poison rates.}
\label{tab:benign_acc}
\begin{adjustbox}{max width=.4\textwidth}
\begin{tabular}{|c|c|c|}
\hline
\textbf{Model} & \textbf{Poison Rate(\%)} & \textbf{Benign Accuracy (\%)} \\
\hline
Clean  & - & 69.71 \\
Backdoored  & 0.16 & 69.72 $\pm$ 0.30 \\
Backdoored  & 0.39 & 69.66 $\pm$ 0.20 \\
Backdoored  & 0.62 & 69.70 $\pm$ 0.07 \\
\hline
\end{tabular}
\end{adjustbox}
\end{table}

\subsection{Transferability}\label{sec: Transferability}
Table~\ref{tab:asr_ba_comparison} demonstrates the performance of \verb"IU" on ResNet-50. The results show that the attack remains effective on a stronger backbone, achieving high ASR across different poison rates. Moreover, due to the superior representation capability of ResNet-50, both BA and ASR are improved compared to the results obtained on ResNet-18 under the same settings.

\begin{table}[h]
\centering
\caption{Comparison of ASR and BA on ResNet-50}
\label{tab:asr_ba_comparison}
\begin{adjustbox}{max width=.3\textwidth}
\begin{tabular}{|c|c|c|}
\hline
\textbf{Poison Rate (\%)} & \textbf{ASR (\%)} & \textbf{BA (\%)} \\
\hline
0.16 & 79.35 & 74.60 \\
0.39 & 90.66 & 75.17 \\
0.62 & 93.75 & 74.84 \\
\hline
\end{tabular}
\end{adjustbox}
\end{table}

Table~\ref{tab:asr_ba_comparison_vit} shows that the ASR of ViT is lower than that of CNN-based models under the same poison rates. This suggests that the backdoor trigger generated from a CNN-based pretrained model transfers well to other CNN architectures, but only partially to the ViT architecture due to the inherent difference in feature representations between CNNs and Vision Transformers. A similar phenomenon has been observed in \cite{10204376,10.1609/aaai.v37i1.25125,Benz2021AdversarialRC}, where the effectiveness of identical attack and defense methods diminishes when transferred across these two architectural families. Nevertheless, even under cross-architecture transfer, \verb"IU" still achieves a strong ASR of 75.4\% at a poison rate of 0.62\%, demonstrating that \verb"IU" maintains considerable effectiveness and robustness across diverse model architectures.

\begin{table}[h]
\centering
\caption{Comparison of ASR and BA on ViT}
\label{tab:asr_ba_comparison_vit}
\begin{adjustbox}{max width=.3\textwidth}
\begin{tabular}{|c|c|c|}
\hline
\textbf{Poison Rate (\%)} & \textbf{ASR (\%)} & \textbf{BA (\%)} \\
\hline
0.16 & 5.10 & 61.62 \\
0.39 & 57.72 & 61.46 \\
0.62 & 75.37 & 61.15 \\
\hline
\end{tabular}
\end{adjustbox}
\end{table}

\subsection{Potential Defense}\label{sec: Potential Defense}
Backdoor defenses fall into two categories: backdoor removal and backdoor detection. The former cleanses the model to ensure any backdoor is eliminated, while the latter only identifies whether a backdoor exists and cannot remove it. Below, we examine existing methods from both categories that apply to \verb"IU".

\paragraph{Backdoor Removal}\label{sec: Backdoor Removal}
Table~\ref{tab:defense_asr} evaluates the robustness of \verb"IU" against three widely-used backdoor removal methods: Fine-Tuning, Fine-Pruning~\cite{liu2018fine}, and NAD~\cite{li2021neural}. We conduct experiments on ResNet-18 using triggers with PSNR = 30 and 1\% of the ImageNet-1K dataset as reserved clean samples, ensuring that the benign accuracy drop remains within 2\%.

As shown in the table~\ref{tab:defense_asr}, \verb"IU" remains resilient against all three removal methods, maintaining a high ASR across different poison rates. Most defenses reduce the ASR by less than 5\%, with only a few cases exceeding a 5\% drop. It demonstrates that \verb"IU" remains largely effective and difficult to mitigate using existing removal methods.

\begin{table}[h]
\centering
\caption{ASR under various defenses and poison rates.}
\label{tab:defense_asr}
\resizebox{\linewidth}{!}{%
\begin{tabular}{|c|c|c|c|}
\hline
\textbf{Defense} & \textbf{Poison Rate(\%)} & \textbf{ASR (\%)} & {$\Delta$ASR(\%)} \\
\hline
\multirow{3}{*}{Fine-Tuning}
& 0.16 & 65.3 & {$\downarrow$ 7.7} \\
& 0.39 & 81.9 & {$\downarrow$ 3.9} \\
& 0.62 & 85.5 & {$\downarrow$ 5.8} \\
\hline
\multirow{3}{*}{Fine-Pruning \textcolor{blue}{{\scriptsize (RAID'18)}}}
& 0.16 & 65.2 & {$\downarrow$ 7.8} \\
& 0.39 & 82.2 & {$\downarrow$ 3.6} \\
& 0.62 & 87.7 & {$\downarrow$ 3.6} \\
\hline
\multirow{3}{*}{NAD \textcolor{blue}{{\scriptsize (ICLR'21)}}}
& 0.16 & 67.1 & {$\downarrow$ 4.9} \\
& 0.39 & 81.1 & {$\downarrow$ 4.7} \\
& 0.62 & 84.3 & {$\downarrow$ 7.0} \\
\hline
\end{tabular}
}
\end{table}

\paragraph{Backdoor Detection}\label{sec: Backdoor Detection} We evaluate \verb"IU" against five representative detection methods: STRIP~\cite{gao2019strip} , SCALE-UP~\cite{guo2023scale}, IBD-PSC~\cite{hou2024ibd}, BARBIE~\cite{DBLP:conf/ndss/ZhangBCM025}, and MM-BD~\cite{10646729}. As shown in Table~\ref{tab:detection}, across all three detection methods and varying poisoning sample sizes, both AUROC and F1-score values remain consistently low. Most AUROC values do not exceed 0.5, except in the case of IBD-PSC, where relatively higher values are observed at poison rates of 0.62\% and 0.39\%. However, when the poison rate is reduced to 0.16\%, the attack become undetectable. 

\begin{table}[h]
\centering
\caption{Detection results under various detection methods.}
\label{tab:detection}
\resizebox{\linewidth}{!}{%
\begin{tabular}{|c|c|c|c|}
\hline
\textbf{Detection} & \textbf{Poison Rate(\%)} & \textbf{AUROC$\downarrow$} & \textbf{F1-score$\downarrow$} \\
\hline
\multirow{3}{*}{STRIP \textcolor{blue}{{\scriptsize (ACSAC'19)}}}
& 0.16 & \textbf{0.31} & \textbf{0.07} \\
& 0.39 & 0.37 & 0.13 \\
& 0.62 & 0.42 & 0.13 \\
\hline
\multirow{3}{*}{SCALE-UP \textcolor{blue}{{\scriptsize (ICLR'23)}}}
& 0.16 & \textbf{0.44} & \textbf{0.02} \\
& 0.39 & 0.46 & 0.08 \\
& 0.62 & 0.45 & 0.05 \\
\hline
\multirow{3}{*}{IBD-PSC \textcolor{blue}{{\scriptsize (ICML'24)}}}
& 0.16 & \textbf{0.27} & \textbf{0.04} \\
& 0.39 & 0.75 & 0.05 \\
& 0.62 & 0.65 & 0.05 \\
\hline
\end{tabular}
}
\end{table}

\begin{table}[h]
\centering
\caption{Detection results under BARBIE and MMBD.}
\label{tab:detection2}
\begin{adjustbox}{max width=.45\textwidth}
\begin{tabular}{|c|c|c|c|}
\hline
\textbf{Poison Rate(\%)} & \textbf{$0.16\%$} & \textbf{$0.39\%$} & \textbf{$0.62\%$} \\
\hline
BARBIE (violations) \textcolor{blue}{{\scriptsize (NDSS'25)}} & 2 & 3 & 4 \\
\hline
MM-BD ($p$-value) \textcolor{blue}{{\scriptsize (IEEE SP'24)}}& 0.0751 & 0.1196 & 0.1286 \\
\hline
\end{tabular}
\end{adjustbox}
\end{table}

For BARBIE, following the original setup, we set the maximum weight of expansion $\omega_{\text{max}}=4.5$ and the upper limit of violations $p=5$, given that only a few clean models were available. For MM-BD, we followed the original configuration with the maximum margin statistic set to $0.05$, so models with a p-value below $0.05$ were classified as backdoored. As shown in Table~\ref{tab:detection2}, none of our backdoored models reached the violation threshold in BARBIE, and all were likewise identified as clean by MM-BD.

These results confirm that our multi-target attack maintains high stealthiness and remains difficult to detect under traditional detection schemes.

\section{Conclusion}
\label{ch:conclusion}
We presented \verb"IU", an imperceptible universal backdoor attack that leverages graph convolutional networks to generate coordinated, class-specific triggers under extremely low poisoning budgets. Through the proposed Trigger Separability Index (TSI), we provided theoretical justification linking feature-space separability to attack success. Experiments on ImageNet-1K demonstrate that IU achieves high ASR and strong stealthiness while evading state-of-the-art defenses. This work highlights the emerging risk of structure-aware, invisible universal backdoors, motivating future research on graph-based defense strategies.

{\small
\bibliographystyle{plain}
\bibliography{ref}

@String(CVPR= {IEEE Conf. Comput. Vis. Pattern Recog.})

@String(ICCV= {Int. Conf. Comput. Vis.})

@String(NIPS= {Adv. Neural Inform. Process. Syst.})

@String(ICLR = {Int. Conf. Learn. Represent.})

@String(AAAI = {AAAI})

@String(CVPR  = {CVPR})

@String(ICCV  = {ICCV})

@String(NIPS  = {NeurIPS})

@String(ICLR  = {ICLR})

@article{gu2017badnets,
  title={Badnets: Identifying vulnerabilities in the machine learning model supply chain},
  author={Gu, Tianyu and Dolan-Gavitt, Brendan and Garg, Siddharth},
  journal={arXiv preprint arXiv:1708.06733},
  year={2017}
}

@inproceedings{doan2021lira,
  title={Lira: Learnable, imperceptible and robust backdoor attacks},
  author={Doan, Khoa and Lao, Yingjie and Zhao, Weijie and Li, Ping},
  booktitle={IEEE/CVF international conference on computer vision (ICCV)},
  pages={11966--11976},
  year={2021}
}

@inproceedings{jiang2023color,
  title={Color backdoor: A robust poisoning attack in color space},
  author={Jiang, Wenbo and Li, Hongwei and Xu, Guowen and Zhang, Tianwei},
  booktitle={IEEE/CVF conference on computer vision and pattern recognition (CVPR)},
  pages={8133--8142},
  year={2023}
}

@article{chen2017targeted,
  title={Targeted backdoor attacks on deep learning systems using data poisoning},
  author={Chen, Xinyun and Liu, Chang and Li, Bo and Lu, Kimberly and Song, Dawn},
  journal={arXiv preprint arXiv:1712.05526},
  year={2017}
}

@inproceedings{nguyen2021wanet,
  title={Wanet--imperceptible warping-based backdoor attack},
  author={Nguyen, Anh and Tran, Anh},
  booktitle={International Conference on Learning Representations (ICLR)},
  year={2021}
}

@inproceedings{cheng2024lotus,
  title={Lotus: Evasive and resilient backdoor attacks through sub-partitioning},
  author={Cheng, Siyuan and Tao, Guanhong and Liu, Yingqi and Shen, Guangyu and An, Shengwei and Feng, Shiwei and Xu, Xiangzhe and Zhang, Kaiyuan and Ma, Shiqing and Zhang, Xiangyu},
  booktitle={IEEE/CVF Conference on Computer Vision and Pattern Recognition (CVPR)},
  year={2024}
}

@inproceedings{
doan2022marksman,
title={Marksman Backdoor: Backdoor Attacks with Arbitrary Target Class},
author={Khoa D Doan and Yingjie Lao and Ping Li},
booktitle={Advances in Neural Information Processing Systems (NeurIPS)},
editor={Alice H. Oh and Alekh Agarwal and Danielle Belgrave and Kyunghyun Cho},
year={2022},
url={https://openreview.net/forum?id=i-k6J4VkCDq}
}

@inproceedings{schneider2023universal,
  title={Universal backdoor attacks},
  author={Schneider, Benjamin and Lukas, Nils and Kerschbaum, Florian},
  booktitle={International Conference on Learning Representations (ICLR)},
  year={2024}
}

@article{turner2019label,
  title={Label-consistent backdoor attacks},
  author={Turner, Alexander and Tsipras, Dimitris and Madry, Aleksander},
  journal={arXiv preprint arXiv:1912.02771},
  year={2019}
}

@inproceedings{li2021invisible,
  title={Invisible backdoor attack with sample-specific triggers},
  author={Li, Yuezun and Li, Yiming and Wu, Baoyuan and Li, Longkang and He, Ran and Lyu, Siwei},
  booktitle={IEEE/CVF international conference on computer vision (ICCV)},
  year={2021}
}

@inproceedings{gao2019strip,
  title={Strip: A defence against trojan attacks on deep neural networks},
  author={Gao, Yansong and Xu, Change and Wang, Derui and Chen, Shiping and Ranasinghe, Damith C and Nepal, Surya},
  booktitle={Annual Computer Security Applications Conference (ACSAC)},
  year={2019}
}

@inproceedings{guo2023scale,
  title={Scale-up: An efficient black-box input-level backdoor detection via analyzing scaled prediction consistency},
  author={Guo, Junfeng and Li, Yiming and Chen, Xun and Guo, Hanqing and Sun, Lichao and Liu, Cong},
  booktitle={International Conference on Learning Representations (ICLR)},
  year={2023}
}

@inproceedings{li2023reconstructive,
  title={Reconstructive neuron pruning for backdoor defense},
  author={Li, Yige and Lyu, Xixiang and Ma, Xingjun and Koren, Nodens and Lyu, Lingjuan and Li, Bo and Jiang, Yu-Gang},
  booktitle={International Conference on Machine Learning (ICML)},
  year={2023}
}

@inproceedings{he2016deep,
  title={Deep residual learning for image recognition},
  author={He, Kaiming and Zhang, Xiangyu and Ren, Shaoqing and Sun, Jian},
  booktitle={IEEE conference on computer vision and pattern recognition (CVPR)},
  year={2016}
}

@inproceedings{bober2023architectural,
  title={Architectural backdoors in neural networks},
  author={Bober-Irizar, Mikel and Shumailov, Ilia and Zhao, Yiren and Mullins, Robert and Papernot, Nicolas},
  booktitle={IEEE/CVF Conference on Computer Vision and Pattern Recognition (CVPR)},
  year={2023}
}

@article{xue2020one,
  title={One-to-N \& N-to-One: Two advanced backdoor attacks against deep learning models},
  author={Xue, Mingfu and He, Can and Wang, Jian and Liu, Weiqiang},
  journal={IEEE Transactions on Dependable and Secure Computing},
  volume={19},
  number={3},
  pages={1562--1578},
  year={2020},
  publisher={IEEE}
}

@inproceedings{10.5555/3327757.3327896,
author = {Tran, Brandon and Li, Jerry and M\k{a}dry, Aleksander},
title = {Spectral signatures in backdoor attacks},
year = {2018},
publisher = {Curran Associates Inc.},
address = {Red Hook, NY, USA},
booktitle = {Advances in Neural Information Processing Systems  (NeurIPS)},
pages = {8011–8021},
numpages = {11},
location = {Montr\'{e}al, Canada},
series = {NIPS'18}
}

@inproceedings{li2021neural,
  title={Neural attention distillation: Erasing backdoor triggers from deep neural networks},
  author={Li, Yige and Lyu, Xixiang and Koren, Nodens and Lyu, Lingjuan and Li, Bo and Ma, Xingjun},
  booktitle={International Conference on Learning Representations (ICLR)},
  year={2021}
}

@inproceedings{wang2019neural,
  title={Neural cleanse: Identifying and mitigating backdoor attacks in neural networks},
  author={Wang, Bolun and Yao, Yuanshun and Shan, Shawn and Li, Huiying and Viswanath, Bimal and Zheng, Haitao and Zhao, Ben Y},
  booktitle={IEEE symposium on security and privacy (SP)},
  year={2019}
}

@inproceedings{liu2018fine,
  title={Fine-pruning: Defending against backdooring attacks on deep neural networks},
  author={Liu, Kang and Dolan-Gavitt, Brendan and Garg, Siddharth},
  booktitle={International symposium on research in attacks, intrusions, and defenses (RAID)},
  year={2018}
}

@inproceedings{hou2024ibd,
  title={Ibd-psc: Input-level backdoor detection via parameter-oriented scaling consistency},
  author={Hou, Linshan and Feng, Ruili and Hua, Zhongyun and Luo, Wei and Zhang, Leo Yu and Li, Yiming},
  booktitle={International Conference on Machine Learning (ICML)},
  year={2024}
}

@inproceedings{
kipf2017semisupervised,
title={Semi-Supervised Classification with Graph Convolutional Networks},
author={Thomas N. Kipf and Max Welling},
booktitle={International Conference on Learning Representations (ICLR)},
year={2017}
}

@inproceedings{NEURIPS2024_4ce18228,
 author = {Xia, Jun and Yue, Zhihao and Zhou, Yingbo and Ling, Zhiwei and Shi, Yiyu and Wei, Xian and Chen, Mingsong},
 booktitle = {Advances in Neural Information Processing Systems (NeurIPS)},
 editor = {A. Globerson and L. Mackey and D. Belgrave and A. Fan and U. Paquet and J. Tomczak and C. Zhang},
 pages = {43549--43570},
 publisher = {Curran Associates, Inc.},
 title = {WaveAttack: Asymmetric Frequency Obfuscation-based Backdoor Attacks Against Deep Neural Networks},
 url = {https://proceedings.neurips.cc/paper_files/paper/2024/file/4ce18228ececb78bca04cbce069891b1-Paper-Conference.pdf},
 volume = {37},
 year = {2024}
}

@inproceedings{li2023embarrassingly,
  title={An embarrassingly simple backdoor attack on self-supervised learning},
  author={Li, Changjiang and Pang, Ren and Xi, Zhaohan and Du, Tianyu and Ji, Shouling and Yao, Yuan and Wang, Ting},
  booktitle={IEEE/CVF International Conference on Computer Vision (ICCV)},
  year={2023}
}

@article{gong2023b3,
  title={B3: Backdoor attacks against black-box machine learning models},
  author={Gong, Xueluan and Chen, Yanjiao and Yang, Wenbin and Huang, Huayang and Wang, Qian},
  journal={ACM Transactions on Privacy and Security},
  volume={26},
  number={4},
  pages={1--24},
  year={2023},
  publisher={ACM New York, NY}
}

@inproceedings{wei2023aliasing,
  title={Aliasing backdoor attacks on pre-trained models},
  author={Wei, Cheng'an and Lee, Yeonjoon and Chen, Kai and Meng, Guozhu and Lv, Peizhuo},
  booktitle={USENIX Security Symposium},
  year={2023}
}

@inproceedings{cao2024data,
  title={Data free backdoor attacks},
  author={Cao, Bochuan and Jia, Jinyuan and Hu, Chuxuan and Guo, Wenbo and Xiang, Zhen and Chen, Jinghui and Li, Bo and Song, Dawn},
  booktitle={Advances in Neural Information Processing Systems (NeurIPS)},
  year={2024}
}

@inproceedings{zeng2023narcissus,
  title={Narcissus: A practical clean-label backdoor attack with limited information},
  author={Zeng, Yi and Pan, Minzhou and Just, Hoang Anh and Lyu, Lingjuan and Qiu, Meikang and Jia, Ruoxi},
  booktitle={ACM SIGSAC Conference on Computer and Communications Security (CCS)},
  year={2023}
}

@inproceedings{10.5555/3327345.3327509,
author = {Shafahi, Ali and Huang, W. Ronny and Najibi, Mahyar and Suciu, Octavian and Studer, Christoph and Dumitras, Tudor and Goldstein, Tom},
title = {Poison frogs! targeted clean-label poisoning attacks on neural networks},
year = {2018},
booktitle = {Advances in Neural Information Processing Systems (NeurIPS)}
}

@inproceedings{sun2025peftguard,
  title={PEFTGuard: detecting backdoor attacks against parameter-efficient fine-tuning},
  author={Sun, Zhen and Cong, Tianshuo and Liu, Yule and Lin, Chenhao and He, Xinlei and Chen, Rongmao and Han, Xingshuo and Huang, Xinyi},
  booktitle={IEEE Symposium on Security and Privacy (SP)},
  year={2025}
}

@article{xiang2021reverse,
  title={Reverse engineering imperceptible backdoor attacks on deep neural networks for detection and training set cleansing},
  author={Xiang, Zhen and Miller, David J and Kesidis, George},
  journal={Computers \& Security},
  volume={106},
  pages={102280},
  year={2021},
  publisher={Elsevier}
}

@inproceedings{deng2009imagenet,
  title={Imagenet: A large-scale hierarchical image database},
  author={Deng, Jia and Dong, Wei and Socher, Richard and Li, Li-Jia and Li, Kai and Fei-Fei, Li},
  booktitle={IEEE conference on computer vision and pattern recognition (CVPR)},
  year={2009}
}

@article{lecun2002gradient,
  title={Gradient-based learning applied to document recognition},
  author={LeCun, Yann and Bottou, L{\'e}on and Bengio, Yoshua and Haffner, Patrick},
  journal={Proceedings of the IEEE},
  volume={86},
  number={11},
  pages={2278--2324},
  year={2002},
  publisher={Ieee}
}

@article{chen2024invisible,
  title={An invisible backdoor attack based on semantic feature},
  author={Chen, Yangming},
  journal={arXiv preprint arXiv:2405.11551},
  year={2024}
}

@inproceedings{zhang2025invisible,
  title={Invisible Backdoor Attack against Self-supervised Learning},
  author={Zhang, Hanrong and Wang, Zhenting and Li, Boheng and Lin, Fulin and Han, Tingxu and Jin, Mingyu and Zhan, Chenlu and Du, Mengnan and Wang, Hongwei and Ma, Shiqing},
  booktitle={IEEE/CVF Conference of the Computer Vision and Pattern Recognition (CVPR)},
  pages={25790--25801},
  year={2025}
}

@article{wang2004image,
  title={Image quality assessment: from error visibility to structural similarity},
  author={Wang, Zhou and Bovik, Alan C and Sheikh, Hamid R and Simoncelli, Eero P},
  journal={IEEE transactions on image processing},
  volume={13},
  number={4},
  pages={600--612},
  year={2004},
  publisher={IEEE}
}

@inproceedings{zhang2018unreasonable,
  title={The unreasonable effectiveness of deep features as a perceptual metric},
  author={Zhang, Richard and Isola, Phillip and Efros, Alexei A and Shechtman, Eli and Wang, Oliver},
  booktitle={IEEE conference on computer vision and pattern recognition (CVPR)},
  year={2018}
}

@inproceedings{NEURIPS2019_9d63484a,
 author = {Yun, Seongjun and Jeong, Minbyul and Kim, Raehyun and Kang, Jaewoo and Kim, Hyunwoo J},
 booktitle = {Advances in Neural Information Processing Systems (NeurIPS)},
 editor = {H. Wallach and H. Larochelle and A. Beygelzimer and F. d\textquotesingle Alch\'{e}-Buc and E. Fox and R. Garnett},
 pages = {},
 publisher = {Curran Associates, Inc.},
 title = {Graph Transformer Networks},
 url = {https://proceedings.neurips.cc/paper_files/paper/2019/file/9d63484abb477c97640154d40595a3bb-Paper.pdf},
 volume = {32},
 year = {2019}
}

@inproceedings{gat,
title={Graph Attention Networks},
author={Petar Veličković and Guillem Cucurull and Arantxa Casanova and Adriana Romero and Pietro Liò and Yoshua Bengio},
booktitle={International Conference on Learning Representations (ICLR)},
year={2018}
}

@inproceedings{DBLP:conf/ndss/ZhangBCM025,
  author={Hanlei Zhang and Yijie Bai and Yanjiao Chen and Zhongming Ma and Wenyuan Xu},
  title={BARBIE: Robust Backdoor Detection Based on Latent Separability},
  year={2025},
  booktitle={Network and Distributed System Security Symposium (NDSS)}
}

@INPROCEEDINGS{10646729,
  author={Wang, Hang and Xiang, Zhen and Miller, David J. and Kesidis, George},
  booktitle={IEEE Symposium on Security and Privacy (SP)}, 
  title={MM-BD: Post-Training Detection of Backdoor Attacks with Arbitrary Backdoor Pattern Types Using a Maximum Margin Statistic}, 
  year={2024}
}

@INPROCEEDINGS{10204376,
  author={Yuan, Zenghui and Zhou, Pan and Zou, Kai and Cheng, Yu},
  booktitle={2023 IEEE/CVF Conference on Computer Vision and Pattern Recognition (CVPR)}, 
  title={You Are Catching My Attention: Are Vision Transformers Bad Learners under Backdoor Attacks?}, 
  year={2023},
  volume={},
  number={},
  pages={24605-24615},
  doi={10.1109/CVPR52729.2023.02357}
}

@inproceedings{10.1609/aaai.v37i1.25125,
author = {Doan, Khoa D. and Lao, Yingjie and Yang, Peng and Li, Ping},
title = {Defending backdoor attacks on vision transformer via patch processing},
year = {2023},
isbn = {978-1-57735-880-0},
publisher = {AAAI Press},
url = {https://doi.org/10.1609/aaai.v37i1.25125},
doi = {10.1609/aaai.v37i1.25125},
booktitle = {Proceedings of the Thirty-Seventh AAAI Conference on Artificial Intelligence and Thirty-Fifth Conference on Innovative Applications of Artificial Intelligence and Thirteenth Symposium on Educational Advances in Artificial Intelligence},
articleno = {56},
numpages = {10},
series = {AAAI'23/IAAI'23/EAAI'23}
}

@inproceedings{Benz2021AdversarialRC,
  title={Adversarial Robustness Comparison of Vision Transformer and MLP-Mixer to CNNs},
  author={Philipp Benz},
  booktitle={British Machine Vision Conference},
  year={2021},
  url={https://api.semanticscholar.org/CorpusID:235806135}
}
}

\onecolumn
\section{Enlarged images from Figure~\ref{fig:trigger_visuals}}
\label{ch:appendix}
\begin{figure*}[ht]
    \centering
    \includegraphics[width=\textwidth]{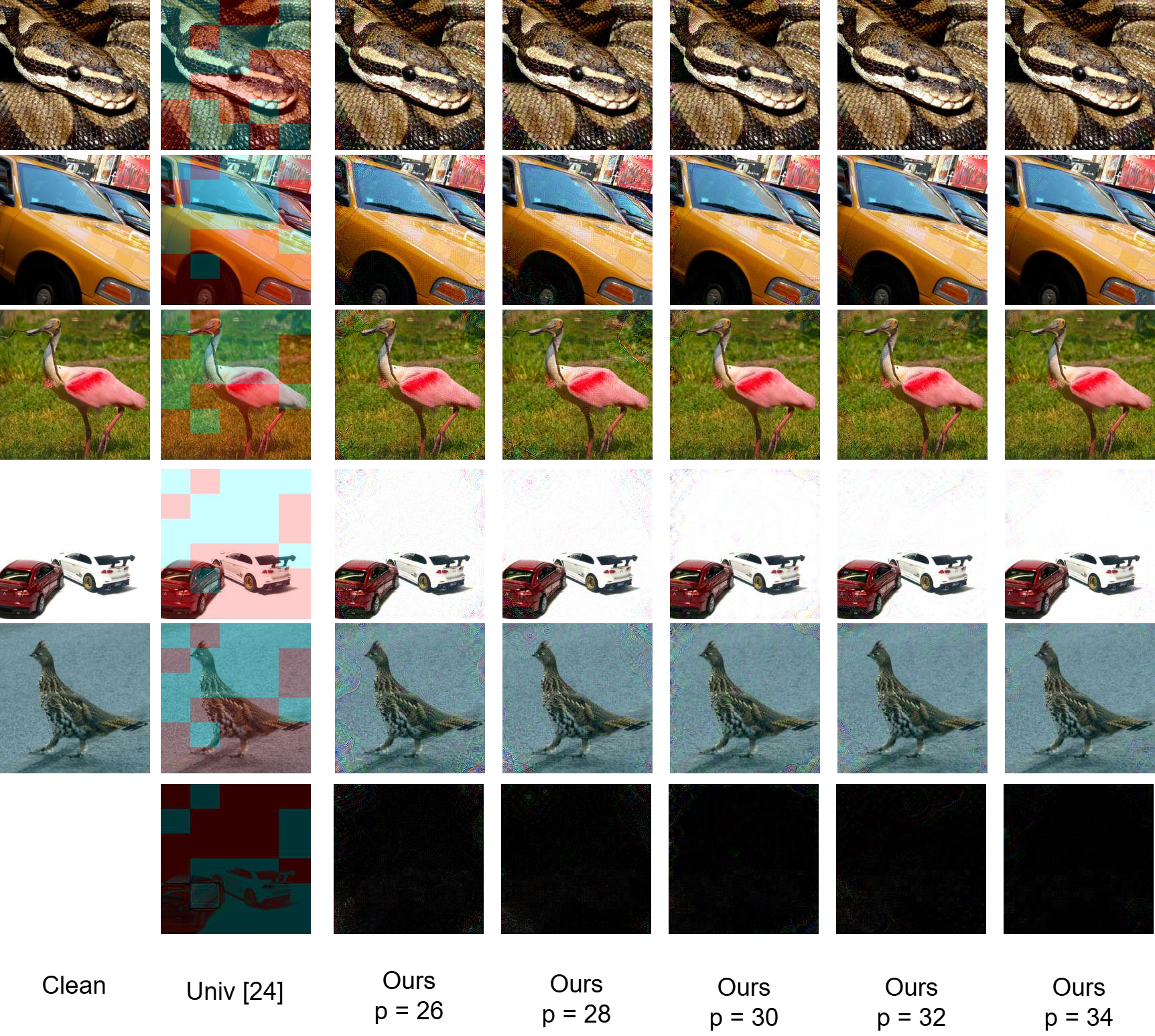}
    \caption{Comparison between different triggers in original size.}
    \label{fig:appendix_image}
\end{figure*}

Due to space constraints, the images in Figure~\ref{fig:trigger_visuals} are small, which may give the impression that the trigger is invisible when in fact the trigger is simply too small for the human eye to detect at that scale. To clarify this point, we provide enlarged versions of the images from Figure~\ref{fig:trigger_visuals} in Figure~\ref{fig:appendix_image}; these larger images demonstrate that the trigger remains imperceptible to human observers.

\section{Notation Table}
Table~\ref{tab:notations} summarizes the notations used throughout this paper.

\begin{table}[htbp]
\centering
\scriptsize
\caption{List of Notations and Parameters Used in This Thesis}
\label{tab:notations}
\renewcommand{\arraystretch}{1.2} 
\begin{tabularx}{\textwidth}{>{\centering\arraybackslash}m{2.5cm} X}
\hline
\textbf{Symbol} & \textbf{Description} \\
\hline

$n$ & The length of latent code after linear discriminant analysis \\
$t$ & Distance threshold for forming an edge — if the sum of absolute differences between two nodes is less than this value, an edge is created \\
$d_{ij}$ & Distance between the features of node $i$ and $j$, computed as the sum of absolute differences \\
$\text{Weight}_{\min}$ & Minimum weight in the graph \\
$T$ & Set of triggers \\
$N$ & Number of target classes \\
$C,H,W$ & Channels, height, and width of the trigger \\
$T_{y'}$ & Trigger corresponding to target class $y'$ \\
$D$ & Clean dataset \\
$D_{\text{sample}}$ & Subset of $D$ used to generate latent representations \\
$D_Z$ & Latent representations extracted from $D_{\text{sample}}$ \\
$\hat{D}_Z$ & Latent representations after LDA compression \\
$M$ & Set of class-wise mean latent representations \\
$c$ & Centroid of all class means \\
$\Delta$ & Difference vector between a class mean and the centroid \\
$b_i$ & $i$-th element of a binary code vector \\
$B_y$ & Binary code vector for class $y$ \\
$D_{\text{poison}}$ & Poisoned dataset containing trigger-injected samples \\
$D_{\text{clean}}$ & Clean dataset before poisoning \\
$P$ & Set of poisoned samples \\
$Y$ & Set of target class labels \\
$y'$ & Target class label \\
$y_t$ & Current target class in poisoning \\
$X$ & Clean image from dataset \\
$X^{\text{poison}}_{y'}$ & Poisoned image corresponding to target class $y'$ \\
$\oplus$ & Element-wise (pixel-wise) addition between the image and the trigger \\
$f_{\theta}$ & Victim model with parameters $\theta$ \\
$f_{\theta'}$ & Pretrained model used to extract latent codes \\
$G$ & Graph constructed from class similarity relations \\
$\mathcal{H}$ & Graph built from binary latent codes \\
$\mathcal{V}$ & Set of nodes in $\mathcal{H}$ \\
$\mathcal{E}$ & Set of edges in $\mathcal{H}$ \\
$W_{ij}$ & Edge weight between class $i$ and $j$ \\
$\mathcal{G}$ & Graph Convolutional Network model \\
$\theta_{\mathcal{G}}$ & Parameters of GCN model $\mathcal{G}$ \\
$\gamma$ & Learning rate for $\mathcal{G}$ training \\
$\delta$ & Gradient of loss w.r.t. $\theta_{\mathcal{G}}$ \\
$\mathcal{L}$ & Total loss function \\
$\mathcal{L}_{\text{stealth}}$ & Stealth loss controlling the stealthiness of trigger \\
$\mathcal{L}_{\text{stealth}}^i$ & Stealth loss for class $i$ \\
$p$ & PSNR threshold controlling trigger imperceptibility (in training) \\
$p$ (poisoning) & Number of poisoned images to insert (in Dataset Poisoning) \\
$\mathcal{L}_{\text{attack}}$ & Attack loss controlling ASR of trigger \\
$\mathcal{L}_{\text{attack}}^i$ & Attack loss for class $i$ \\
$\text{CE}(,)$ & Cross-entropy loss \\
$\beta$ & Trade-off coefficient between stealthiness and attack success rate \\
$t_{y_t}$ & Trigger corresponding to target class $y_t$ \\
$\hat{x}$ & Image $x$ with trigger $t_{y_t}$ applied \\
$\text{ASR}_{y'}$ & Attack Success Rate on target class $y'$ \\
ASR & Average Attack Success Rate over all target classes \\
BA & Benign Accuracy \\
$T_{y'}$ & The trigger corresponding to target class $y'$ \\
\hline
\end{tabularx}
\end{table}

\section{Why TSI? Relation to Classical Separability Measures and to ASR}

Following our notation, let $f(x)=\operatorname{softmax}(W\cdot \phi(x))$ with last-layer weights $W=[w_1,\dots,w_K]^\top$ and feature $\phi(x)\in\mathbb{R}^d$. For a target class $y'$, define the average feature displacement caused by its trigger $T_{y'}$ as
\[
v_{y'} \triangleq \mathbb{E}_{x\sim D}\big[\phi(x+T_{y'})-\phi(x)\big] \tag{Eqs.~\ref{eq: 7} and \ref{eq: 8} in our paper}
\]
and the per-class logit gap after triggering as
\[
\Delta_{y',k} \triangleq (w_{y'}-w_k)^\top v_{y'},\quad k\neq y'. \tag{Eq.~\ref{eq: 8} in our paper}
\]
Larger mean and smaller variance of $\{\Delta_{y',k}\}_{k\neq y'}$ intuitively imply higher attack success.\footnote{See our Theoretical Justification where we motivate this view.}
We introduced the \emph{Trigger Separability Index (TSI)}:
\[
\operatorname{TSI}(y') \;\triangleq\; \frac{\mathbb{E}_{k\neq y'}[\Delta_{y',k}]}
{\sqrt{\mathrm{Var}_{k\neq y'}(\Delta_{y',k})+\varepsilon}}, \quad \varepsilon>0. \tag{Eq.~\ref{eq: 9} in our paper}
\]

\paragraph{TSI vs.\ Fisher discriminant ratio (FDR).}
Classical Fisher ratio measures \emph{between-class mean separation} vs.\ \emph{within-class variance}. 
For a given pair $(y',k)$, a 1-D Fisher ratio along the direction $(w_{y'}-w_k)$ can be written as
\[
\mathrm{FDR}_{y',k} \;=\; 
\frac{\big(\mu_{y',k}^{\text{trig}}-\mu_{y',k}^{\text{base}}\big)^2}{\sigma^2_{y',k,\text{trig}}+\sigma^2_{y',k,\text{base}}},
\]
where $\mu^{\text{trig}}_{y',k}$ (resp.\ $\mu^{\text{base}}_{y',k}$) and $\sigma^2$ are the mean and variance of $(w_{y'}-w_k)^\top\phi(\cdot)$ under triggered (resp.\ clean) inputs.
While $\mathrm{FDR}_{y',k}$ is meaningful \emph{pairwise}, the universal backdoor setting requires \emph{consistency across \emph{all} $k\neq y'$}.
In contrast, TSI \emph{aggregates} the cross-class signal-to-noise property by
averaging the pairwise gaps $\Delta_{y',k}$ in the numerator and normalizing by their \emph{cross-class} variance in the denominator.
Hence, TSI directly captures the \emph{class-wise coordination} needed by universal attacks, which FDR does not measure unless one introduces an extra aggregation heuristic (e.g., minimum/mean across $k$), whose statistical meaning for top-1 decision is unclear.

\paragraph{TSI vs.\ margin-based measures.}
A natural margin surrogate for success into $y'$ is 
\[
m \;\triangleq\; \min_{k\neq y'} \Delta_{y',k}.
\]
Maximizing $m$ promotes success but is hard to estimate due to the $\min$ over $K{-}1$ classes.
Under Gaussian/sub-Gaussian assumptions for $\{\Delta_{y',k}\}_{k\neq y'}$ with common mean $\mu$ and variance proxy $\sigma^2$, classical bounds on extrema imply 
\[
\mathbb{E}[m] \;\gtrsim\; \mu - \sigma\sqrt{2\log(K-1)}.
\]
Thus, requiring $\operatorname{TSI}=\mu/\sigma \gtrsim \sqrt{2\log(K-1)}$ already guarantees a positive expected margin. 
TSI therefore serves as a \emph{computationally stable}, \emph{aggregation-aware} proxy for the intractable minimum-margin.

\paragraph{Assumptions.}
We make the following standard assumptions for concentration: 
(A1) linear last layer (already satisfied in our model); 
(A2) for a uniformly random non-target class $K$ and triggered input distribution, $\Delta \equiv \Delta_{y',K}$ is sub-Gaussian with mean $\mu=\mathbb{E}[\Delta]$ and variance proxy $\sigma^2=\mathrm{Var}(\Delta)$; 
(A3) across $k$, the tails of $\Delta_{y',k}$ are uniformly dominated by the same sub-Gaussian proxy $\sigma^2$.

\paragraph{From TSI to ASR: tail bounds.}
We connect TSI to the attack success rate (ASR), i.e., the probability that $y'$ outruns all $k\neq y'$:
\[
\mathrm{ASR}(y') \;=\; \Pr\big(\Delta_{y',k}>0,\ \forall k\neq y'\big).
\]

\begin{proposition}[Per-class tail and union bound]
\label{prop:perclass}
Under (A1)--(A3), for each $k\neq y'$,
\(
\Pr(\Delta_{y',k}\le 0)\le \exp\!\big(-\mu^2/(2\sigma^2)\big).
\)
Consequently,
\[
\mathrm{ASR}(y') 
\;\ge\; 1 - (K-1)\exp\!\Big(-\tfrac{1}{2}\operatorname{TSI}(y')^2\Big).
\]
\end{proposition}

\noindent\emph{Sketch.}
For a sub-Gaussian variable $\Delta$ with mean $\mu>0$, the one-sided Chernoff/Cantelli bound gives $\Pr(\Delta\le 0)\le \exp(-\mu^2/(2\sigma^2))$. Union bound over $K{-}1$ competitors yields the result.

\begin{corollary}[TSI threshold for a target success level]
\label{cor:threshold}
To guarantee $\mathrm{ASR}(y')\ge 1-\delta$, it suffices that
\[
\operatorname{TSI}(y') \;\ge\; \sqrt{2\log\big((K-1)/\delta\big)}.
\]
\end{corollary}

\paragraph{From TSI to expected margin.}
Let $M=\min_{k\neq y'}\Delta_{y',k}$.
If $\{\Delta_{y',k}\}$ are i.i.d.\ sub-Gaussian with mean $\mu$ and proxy $\sigma^2$, then
\[
\mathbb{E}[M] 
\;\ge\; \mu - \sigma\sqrt{2\log(K-1)} 
\;=\; \sigma\!\left(\operatorname{TSI}(y') - \sqrt{2\log(K-1)}\right).
\]
Hence $\operatorname{TSI}>\sqrt{2\log(K-1)}$ implies a \emph{positive expected margin}, consistent with Proposition~\ref{prop:perclass}.

\paragraph{Why TSI is necessary in the universal regime.}
Universal backdoors simultaneously enforce \emph{cross-class} displacement in the common direction favoring $y'$. 
FDR captures \emph{pairwise} separability, while margin directly targets the decision rule but is unstable to estimate in high class-count settings. 
TSI, by design, summarizes the \emph{mean-vs-variance} trade-off of $\{\Delta_{y',k}\}_{k\neq y'}$ across all competitors, mirroring a \emph{signal-to-noise} ratio at the class-competition level.
This is precisely what the top-1 decision depends on when triggers must generalize consistently across many source classes.

\section{Comparison of Different GNN Models}\label{sec: Comparison of Different GNN Models}
Table~\ref{tab:gat_vs_gt} presents the comparison between GAT~\cite{gat} and Graph Transformer~\cite{NEURIPS2019_9d63484a} under a poison rate of 0.62\%. Although both models achieve similar BA (69.12\% for GAT and 69.14\% for Graph Transformer), there is a significant difference in ASR. Specifically, GAT reaches 27.05\% ASR, while the Graph Transformer achieves only 0.08\%. This phenomenon arises because the graph structure in these models does not explicitly define edge weights; instead, the relationships are learned implicitly during training. However, the training process does not optimize the loss function with respect to the correlation among triggers. As a result, the ASR observed during GNN training cannot effectively transfer to real poisoned data, leading to poor performance in actual backdoor scenarios. The results in Table~\ref{tab:gat_vs_gt} justifies our design choice in using GCN in \verb"IU".

\begin{table}[h]
\centering
\caption{Comparison of ASR(\%) and BA(\%) between GAT, Graph Transformer and GCN under 0.62\% poison rate.}
\label{tab:gat_vs_gt}
\begin{tabular}{|c|c|c|c|c|c|c|}
\hline
\multirow{2}{*}{\textbf{Poison Rate (\%)}} & \multicolumn{2}{c|}{\textbf{GAT}} & \multicolumn{2}{c|}{\textbf{Graph Transformer}} & \multicolumn{2}{c|}{\textbf{GCN}} \\
\cline{2-7}
 & \textbf{ASR(\%)} & \textbf{BA(\%)} & \textbf{ASR(\%)} & \textbf{BA(\%)} & \textbf{ASR(\%)} & \textbf{BA(\%)} \\
\hline
0.62 & 27.05 & 69.12 & 0.08 & 69.14 & 91.30 & 69.80 \\
\hline
\end{tabular}
\end{table}

\section{Computation Time Comparison}\label{sec: Computation Time Comparison}
Table~\ref{tab:gcn_time} presents the time comparison between GCN training and inference. The results in Table~\ref{tab:gcn_time} were derived on Intel Xeon Silver 4309Y, 64GB RAM, and NVIDIA H100. As shown, the training process requires a longer duration (1:44:18) due to the complexity of learning graph representations. In contrast, the inference time is efficient, requiring only 0.179 seconds. This demonstrates that while training GCNs is computationally expensive, once trained, the model can perform inference rapidly, making it practical for real-world deployment.

\begin{table}[h]
\centering
\caption{Training and inference time of GCN.}
\label{tab:gcn_time}
\begin{tabular}{|c|c|}
\hline
\textbf{Process} & \textbf{Time} \\
\hline
GCN Training & 1:44:18 \\
GCN Inference & 0.179s \\
\hline
\end{tabular}
\end{table}

\section{Attack under CIFAR-10 dataset}\label{sec: Attack under CIFAR-10 dataset}
In the experiments on CIFAR-10, \verb"IU" exhibits relatively poor performance. The GCN-generated triggers are designed to enhance each other’s effectiveness, thereby achieving a higher ASR under low poison rates. However, the CIFAR-10 dataset contains only ten classes, which is insufficient for enabling strong mutual reinforcement among the triggers. As shown in Table~\ref{tab:cifar10_results}, even at a poison rate of 0.62\%, the ASR only reaches 72.0\%, and further decreases to 35.8\% and 23.3\% when the poison rates drop to 0.39\% and 0.16\%, respectively. Consequently, the expected improvement in ASR cannot be fully realized, leading to relatively low overall ASR.

\begin{table}[h]
\centering
\caption{Results on CIFAR-10: ASR and BA under different poison rates.}
\label{tab:cifar10_results}
\begin{tabular}{|c|c|c|}
\hline
\textbf{Poison Rate (\%)} & \textbf{ASR(\%)} & \textbf{BA(\%)} \\
\hline
0.16 & 23.3 & 87.0 \\
0.39 & 35.8 & 87.9 \\
0.62 & 72.0 & 87.9 \\
\hline
\end{tabular}
\end{table}

\section{Performance under different $t$ threshold}\label{sec: Performance under different $t$ threshold}

The experimental results in Table~\ref{tab:psnr_results} demonstrate the impact of different threshold values $t$ on ASR and BA. When $t = 0$, the graph contains no edges between nodes, resulting in no correlation among triggers. Consequently, the triggers cannot reinforce each other, making it difficult to launch effective attacks under low-visibility conditions, which leads to a significantly reduced ASR. In contrast, when $t = 5$, a moderate level of connectivity allows triggers to effectively propagate and amplify the attack signal, achieving the highest ASR across all parameter settings and demonstrating optimal attack efficiency. However, when $t = 10$, the excessive connections among triggers cause interference effects, reducing the consistency of the attack and leading to lower ASR compared to $t = 5$. It is worth noting that BA remains consistently around 69\% across all $t$ values, indicating that the attack primarily affects the target classes while having limited influence on the overall model accuracy. Overall, the results in Table~\ref{tab:psnr_results} justify our design choice in using $t=5$ in \verb"IU".

\begin{table}[h]
\centering
\caption{ASR and BA under different $t$ thresholds with poison rate 0.062\%.}
\label{tab:psnr_results}
\begin{tabular}{|c|c|c|c|c|c|c|c|c|c|c|}
\hline
\multirow{2}{*}{\textbf{$t$}} & \multicolumn{2}{c|}{\textbf{$p=26$}} & \multicolumn{2}{c|}{\textbf{$p=28$}} & \multicolumn{2}{c|}{\textbf{$p=30$}} & \multicolumn{2}{c|}{\textbf{$p=32$}} & \multicolumn{2}{c|}{\textbf{$p=34$}} \\
\cline{2-11}
 & \textbf{ASR(\%)} & \textbf{BA(\%)} & \textbf{ASR(\%)} & \textbf{BA(\%)} & \textbf{ASR(\%)} & \textbf{BA(\%)} & \textbf{ASR(\%)} & \textbf{BA(\%)} & \textbf{ASR(\%)} & \textbf{BA(\%)} \\
\hline
0  & 83.9 & 69.2 & 62.6 & 69.6 & 15.13 & \textbf{69.9} & 1.7 & 69.7 & 1.1 & 69.4 \\
5  & \textbf{93.1} & 69.7 & \textbf{92.0} & 69.7 & \textbf{91.3} & 69.8 & \textbf{83.4} & 69.6 & \textbf{82.0} & 69.7 \\ 
10 & 82.9 & \textbf{69.8} & 81.5 & \textbf{69.9} & 77.6 & 69.7 & 72.5 & \textbf{69.9} & 72.0 & \textbf{69.6} \\
\hline
\end{tabular}
\end{table}

\section{Comparison of TSI under different $t$ threshold}\label{sec: Comparison of TSI under different $t$ threshold}

The results in Table~\ref{tab:tsi_results} illustrate the effect of different $t$ thresholds on the Trigger Separability Index (TSI). When $t = 0$, the generated triggers fail to induce effective attacks, resulting in very low TSI values across all parameter settings. As $t$ increases to 5, the enhanced connectivity among triggers enables successful attacks, which leads to a significant increase in TSI and achieves the highest performance. However, when $t$ is further increased to 10, the excessive influence among triggers introduces interference effects, thereby reducing the stability of the attack and causing TSI to decrease compared to the $t = 5$ case.

\begin{table}[h]
\centering
\caption{TSI under different $t$ thresholds with poison rate 0.62\%.}
\label{tab:tsi_results}
\begin{tabular}{|c|c|c|c|c|c|}
\hline
\textbf{$t$} & \textbf{$p=26$} & \textbf{$p=28$} & \textbf{$p=30$} & \textbf{$p=32$} & \textbf{$p=34$} \\
\hline
0  & 4.8740  & 4.2657 & 2.6903 & 1.203 & 0.4818  \\
5  & \textbf{6.4991} & \textbf{6.3601} & \textbf{6.1893} & \textbf{6.4829} & \textbf{6.2070} \\ 
10 & 5.8762  & 5.8222  & 3.4219  & 5.5263  & 5.3034  \\
\hline
\end{tabular}
\end{table}

\section{Comparison of Attack Success Rates Across Different Classes}\label{sec: Comparison of Attack Success Rates Across Different Classes}
\begin{figure}[h]
\centering
\includegraphics[width=1.0\linewidth]{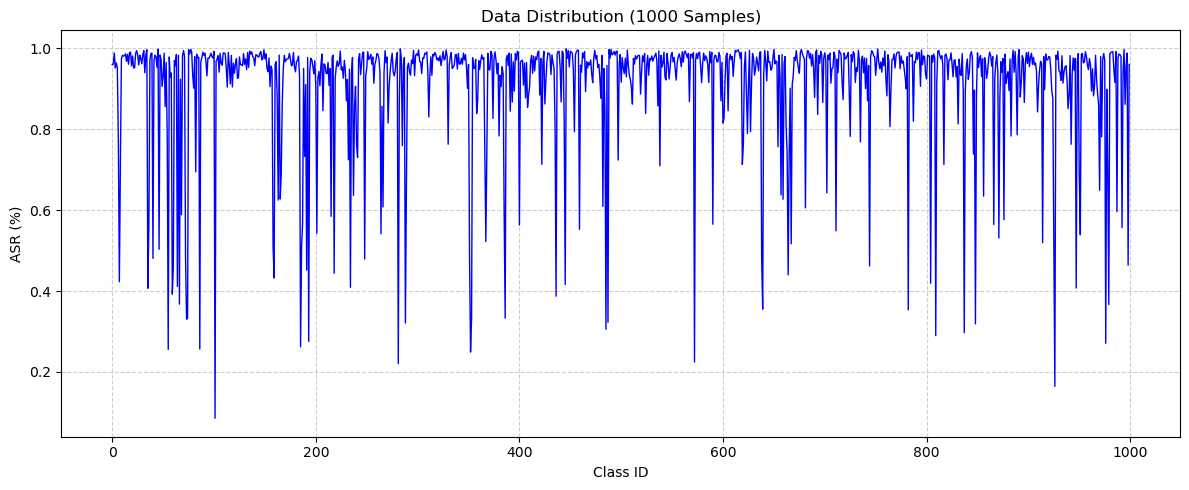}
\caption{Comparison of ASR across different classes}
\label{fig:ASR_list}
\end{figure}

The figure~\ref{fig:ASR_list} and table~\ref{tab:asr_distribution} shows the distribution of ASR across different target classes. Although the results are not perfectly uniform across all classes, but the ASR exceeds 0.9 for most classes (around 800 classes), while fewer than 50 classes have an ASR below 0.5.

\begin{table}[h]
\centering
\caption{Distribution of ASR across 1000 target classes.}
\label{tab:asr_distribution}
\begin{tabular}{|c|c|}
\hline
\textbf{ASR Range (\%)} & \textbf{Count} \\
\hline
$\geq$ 95 & 643 \\
90 -- 95 & 171 \\
85 -- 90 & 51 \\
80 -- 85 & 19 \\
75 -- 80 & 17 \\
70 -- 75 & 15 \\
65 -- 70 & 4 \\
60 -- 65 & 11 \\
55 -- 60 & 12 \\
50 -- 55 & 12 \\
$<$ 50 & 45 \\
\hline
\textbf{Total} & \textbf{1000} \\
\hline
\end{tabular}
\end{table}

\section{Comparison of ASR and BA under different n setting}
As shown in Table \ref{tab:n_ASR}, when the value of $n$ increases, the model can effectively avoid the issue where samples from different classes are projected into the same latent space after LDA dimensionality reduction. A larger $n$ helps maintain better feature separability between the trigger patterns and target classes, resulting in a higher ASR.

\begin{table}[h]
\centering
\caption{Comparison of ASR and BA under different n setting with poison rate 0.62\%}
\label{tab:n_ASR}
\begin{tabular}{|c|c|c|}
\hline
\textbf{$n$} & \textbf{ASR(\%)} & \textbf{BA(\%)} \\
\hline
60 & 95.29 & 69.36 \\
50 & 92.70 & 69.80 \\
40 & 83.81 & 69.04 \\
\hline
\end{tabular}
\end{table}

 \newpage

\end{document}